\documentclass[11pt,a4paper]{article}
\usepackage[hyperref]{tto2019}
\usepackage{times}
\usepackage{latexsym}

\usepackage{url}
\usepackage{graphicx}
\usepackage{booktabs}
\newcommand{\cmmnt}[1]{}

\aclfinalcopy 

\title{Robust Information Retrieval for False Claims with Distracting Entities In Fact Extraction and Verification}

\author{\normalsize 
    Mingwen Dong \\
  \texttt{\small mingwd@amazon.com} \\\And
    \normalsize Christos Christodoulopoulos \thanks{Christos Christodoulopoulos is affiliated with Amazon Alexa. Other authors are affiliated with Amazon Web Services, Inc.} \\
  \texttt{\small chrchrs@amazon.co.uk} \\\And
    \normalsize Sheng-Min Shih \\
  \texttt{\small shengms@amazon.com} \\\And
    \normalsize Xiaofei Ma \\
  \texttt{\small xiaofeim@amazon.com}
}

\date{}

\begin{document}
\maketitle
\begin{abstract}
	Accurate evidence retrieval is essential for automated fact checking. Little previous research has focused on the differences between true and false claims and how they affect evidence retrieval. \cmmnt{For example, false claims can more often contain irrelevant and distracting entities than true claims. Consequently, true claims and supporting evidence often have a large lexical and semantic overlap whereas false claims and refuting evidence may only partially overlap. If an information retrieval model always tries to find evidence with largest lexcial and semantic overlap, some irrelevant entities in a false claim my distract the model.}This paper shows that, compared with true claims, false claims more frequently contain irrelevant entities which can distract evidence retrieval model. A BERT-based retrieval model made more mistakes in retrieving refuting evidence for false claims than supporting evidence for true claims. When tested with adversarial false claims (synthetically generated) containing irrelevant entities, the recall of the retrieval model is significantly lower than that for original claims. These results suggest that the vanilla BERT-based retrieval model is not robust to irrelevant entities in the false claims. By augmenting the training data with synthetic false claims containing irrelevant entities, the trained model achieved higher evidence recall, including that of false claims with irrelevant entities. In addition, using separate models to retrieve refuting and supporting evidence and then aggregating them can also increase the evidence recall, including that of false claims with irrelevant entities. These results suggest that we can increase the BERT-based retrieval model's robustness to false claims with irrelevant entities via data augmentation and model ensemble.
\end{abstract}

\section{Introduction}

Automated fact check has received a lot of attention recently and several datasets have been developed to facilitate relevant research \citep{Vlachos2014FactCT, Wang2017LiarLP, Pomerleau2017FakeNews, hanselowski-etal-2018-retrospective}. However, most of these datasets contain limited amount of examples and the human annotation is only at the document level. Without a large amount of sentence level annotations, it is difficult to train latest neural-network based models to detect false claims. The Fact Extraction and VERification (FEVER) task was introduced to solve these data challenges \citep{thorne-etal-2018-fact}. The task is to fact-check sentences (claims) people wrote to be true or false by retrieving evidence sentences from Wikipedia and using them to verify the validity of the claim. True claims are validated by supporting evidence that entails them and false claims are verified by refuting evidence that contradicts them.

A lot of progress has been made on the FEVER task using classical and deep learning methods \citep{yoneda2018ucl, hanselowski2018ukp, TYSS2019AttentiveCheckerAB, soleimani2020bert, hidey2020deseption, Lee2020LanguageMA}. Our paper focuses on evidence retrieval and we review some related progress. \citet{thorne-etal-2018-fact} used TF-IDF for evidence retrieval to show the feasibility and challenge of the task. \citet{yoneda2018ucl} improved the evidence retrieval accuracy by training logistic regression and MLP models using manually engineered features (e.g., token matching between claim and evidence sentence, evidence sentene length). Based on the observation that most claims focus on Wikipedia entities, \citet{hanselowski2018ukp} extracted entities from claims using AllenNLP and then used those entities as queries to retrieve relevant documents leveraging MediaWiki search API \citep{Gardner2018AllenNLPAD}. \citet{soleimani2020bert} combined the method from \citet{hanselowski2018ukp} with BERT-based models and improved the evidence retrieval performance by 2.5\%. Recently, \citet{hidey2020deseption} achieved state-of-the-art results using a combination of BERT models and pointer networks \citep{vinyals2015pointer, devlin-etal-2019-bert}.

\begin{table*}[]
	\centering
	\caption{A false claim with its refuting evidence from FEVER. ``Stan Beeman'' and ``BBC'' are two entities. ``BBC'' is an irrelevant entity that may distract an information retrieval model.}
	\label{figure-FEVER-data-example}
	\begin{tabular}{ll}
		\toprule
		Claim & Stan Beeman is only in shows on BBC. \\
		Evidence & Stan Beeman acts in a US TV series. \\
		Relationship & \textsc{Refuted} \\
		\bottomrule
	\end{tabular}
\end{table*}

Few research has reported separate results for false and true claims, possibly because they treat evidence retrieval for FEVER as a standard information retrieval (IR) task and used the same IR model to retrieve refuting evidence for false claims and supporting evidence for true claims. \citet{Yoneda2018UCLMR} and \citet{Chakrabarty2018RobustDR} reported that the accuracy is lower in detecting false claims than detecting true claims, which suggests that false claims and true claims have different characteristics. One difference we identified is that false claims are much more likely to contain irrelevant or distracting information than true claims. For example, in the false claim ``Stan Beeman is only in shows on BBC'', ``BBC'' is not related to the actual evidence (Table \ref{figure-FEVER-data-example}) and may mislead a retrieval model to find evidence containing both BBC and Stan Beeman. In other words, an IR model may be susceptible to the distraction of the irrelevant entity ``BBC'' if it retrieves sentences with the largest lexical and semantic overlap with the query. In this paper, we show that a vanilla BERT-based retrieval model is not robust to irrelevant entities in false claims by analyzing the distribution of entities' relationship in different types of claims and testing the model with synthetically generated claims containing irrelevant entities. We used two different methods to improve the retrieval model's robustness. The first method is augmenting false claims with irrelevant entities so that the trained model is more robust to them. The second method is retrieving refuting and supporting evidence using separate models and then aggregating them, which allows the model for refuting evidence retrieval to be trained separately and become more robust to irrelevant entities in false claims. Our contributions are:  1) showing that a vanilla BERT-based retrieval model is not robust to irrelevant entities in false claims; 2) showing two effective methods to increase the model robustness, one by using data augmentation and the other by having separate models to retrieve refuting and supporting evidence and then aggregating.

\begin{table*}[]
	\centering
	\caption{Statistics of FEVER dataset \& synthetic refuted claims. \cmmnt{Training set has more supported claims than refuted claims.}}
	\label{table-FEVER-data-stats}
	\begin{tabular}{l l l l l}
		\toprule
		split & \textsc{Supported} & \textsc{Refuted} & NEI & Synthetic \textsc{Refuted}   \\
		\hline
		Training & 80,035 & 29,775 & 35,639 & 7456 \\
		Dev & 3,333 & 3,333 & 3,333 & 1102 \\
		Test & 3,333 & 3,333 & 3,333 & - \\
		Reserved & 6,666 & 6,666 & 6,666 & - \\
		\bottomrule
	\end{tabular}
\end{table*}

\section{FEVER Task \& Models}

FEVER dataset contains false, true, and unverifiable claims modified from sentences in Wikipedia \citep{thorne-etal-2018-fact}.  The task is to classify each claim as \textit{\textsc{Refuted}}, \textit{\textsc{Supported}}, or \textit{\textsc{NotEnoughInfo}} (\textit{NEI}) by retrieving relevant evidence sentences from Wikipedia and classifying the relationship between a claim and corresponding evidence sentences (Table \ref{figure-FEVER-data-example}).

The FEVER task is often decomposed into three sub tasks: document retrieval, sentence selection, and natural language inference. Our paper focuses on the sentence selection and compares the evidence retrieval recall and fact check accuracy of different sentence selection methods while using the same document retrieval and natural language inference models. Because the official FEVER evaluation code typically expects 5 evidence sentences per claim, we focus on the recall to evaluate evidence retrieval (calculated using official code @k=5 for sentence selection, k=20 for document retrieval) \citep{thorne-etal-2018-fact}. We also counted the number of mistakes in retrieving refuting and supporting evidence separately to measure a model's robustness toward different types of claims. A model is said to make one mistake if the top $k$ evidence retrieved do not contain any ground truth evidence (k=5 for sentence selection, k=20 for document retrieval). Lastly, the official FEVER score is used to measure the overall system performance.


\subsection{Document Retrieval}
The document retrieval model selects the top documents that potentially contain evidence sentences from millions of Wikipedia documents.  The top 20 documents are passed to the sentence selection model to retrieve the sentences relevant to a claim. We adopted the method from \citet{hanselowski2018ukp} because it is efficient, achieves greater than 93\% recall (k=20 as in most previous research), and has been adopted by other previous research \citep{Zhou2019GEARGE, soleimani2020bert}. Table \ref{table-document-retrieval} shows that the adopted document retrieval method made more mistakes in retrieving refuting evidence than supporting evidence.

\begin{table}[]
	\centering
	\addtolength{\tabcolsep}{-3pt}
	\caption{Document retrieval results $@k=20$ using method adopted from \citep{hanselowski2018ukp}.}
	\label{table-document-retrieval}
	\begin{tabular}{l l l }
		\toprule
		Recall & Refuted mistakes & Supported mistakes \\
		\midrule
		0.931 & 147 & 82\\
		\bottomrule
	\end{tabular}
\end{table}

\subsection{Sentence Selection}
\label{different_models}

Given the documents from the document retrieval model, the sentence selection model selects refuting and supporting sentences for false and true claims, respectively. We use BERT-based binary classifiers for sentence selection because BERT is a general method and has been widely used in IR, FEVER and other NLP tasks \citep{devlin-etal-2019-bert, Dai_2019, soleimani2020bert}. To show that a fine-tuned BERT-based retrieval model is not robust to irrelevant entities in false claims, we compared multiple binary classification (relevant vs. irrelevant) BERT-models trained using different subsets of training data: both supported and refuted claims (Baseline), only supported claims (SUP), or only refuted claims (REF). We expect that the baseline model will make more mistakes than REF model for refuted claims (refuted mistake) and make more mistakes than SUP model for supported claims (supported mistake). To get the best of SUP and REF models (SR), we aggregated and reranked their results based on the confidence scores. Note that we also tried to combine results from the SUP and REF model based on the rank or a trained linear regression model, but the results were similar to that based on the confidence score. In the end, we used data augmentation and trained another model (DA), in which the original supported and refuted claims is augmented with synthetically generated false claims described in the next part.

\begin{figure*}[h]
	\centering
	\includegraphics[width=0.63\textwidth]{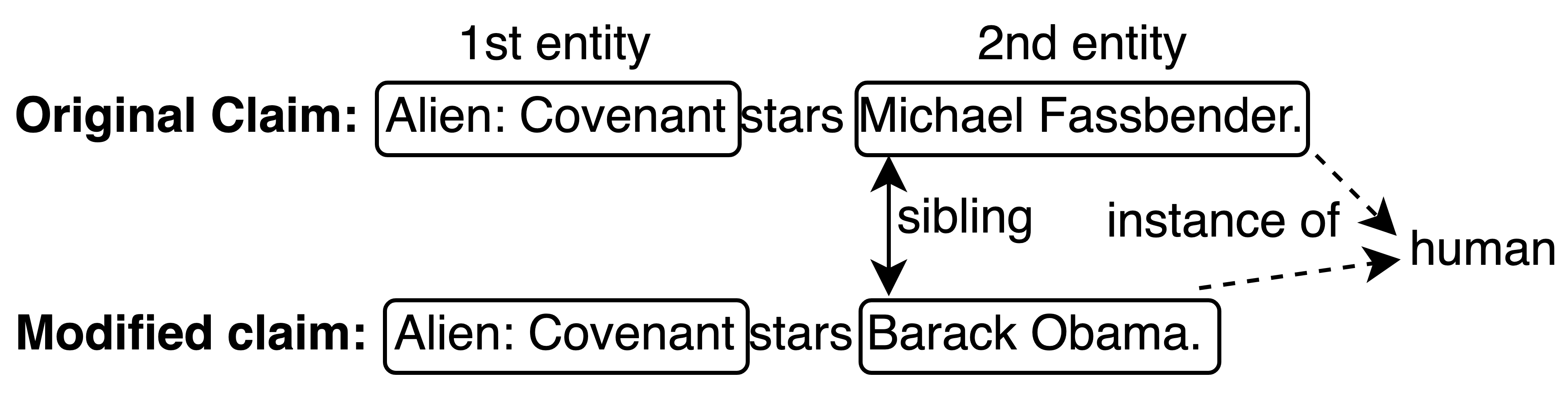}
	\caption{An example false claim generated using entity linking and Wikidata.}
	\label{fig:wikidata_aug_claims}
\end{figure*}

\textbf{Generating False Claims with Distracting Entities.} We hypothesize that retrieving refuting evidence for false claims is more challenging than retrieving supporting evidence for true claims for two reasons. Firstly, false claims more often contain irrelevant entities than true claims and the irrelevant entities can be distracting to the evidence retrieval models which often uses lexical matching as one important feature. Secondly, FEVER data set contains fewer training examples for false (refuted) claims than true (supported) claims (Table \ref{table-FEVER-data-stats}). We programmatically modified the true claims in the training and dev set to create additional false claims with irrelevant and distracting entities. For each true claim, its entities were programmatically recognized and linked to WikiData using the API from \citet{vanHulst:2020:REL}. If two or more entities were identified, the 2nd entity was replaced with one of its sibling entities sampled from WikiData to generate one false claim (Figure \ref{fig:wikidata_aug_claims}). An entity's sibling is any other entities that share the same parent. Notice that not all true claims can have a false claim generated from it because the true claim might not have a second entity with sampled sibling. We chose to replace the 2nd entity to generate false claims because the irrelevant entity of false claims in FEVER tends to be the 2nd one based on our manual check. The evidence of the original claim is treated as the refuting evidence for the generated false claim. In total, 7456 and 1102 false claims were generated from the training and dev set, respectively (Table \ref{table-FEVER-data-stats}). To check the quality of the synthetic false claims, we randomly sampled 100 synthetic claims and manually checked whether they are true or false. Among 100 sampled claims, 73 are false, 22 are hard to determine (probably false but not semantically meaningful), 5 are true. Because synthetic claims are used to train the augmented sentence selection model (DA) which does not distinguish between true and false claims during training, the accidentally generated true and hard to determine claims will not negatively influence the training results.

\textbf{Training.} A binary classifier classifies whether a candidate evidence sentence is relevant to a given claim. The probability of being relevant is used to rank the candidate sentences. All models are fine-tuned to minimize the cross-entropy (negative log-likelihood) loss starting from a pre-trained BERT-base-cased model using the transformers library \citep{Wolf2019HuggingFacesTS}. Each model is fine-tuned for 2 epochs with a learning rate of 2.5e-6. Following previous work, the Wikipedia page title is appended to each candidate sentence during training and inference for pronoun resolution \cite{yoneda2018ucl, soleimani2020bert}.

 
\textbf{Negative Sampling.} For each supported or refuted claim in the training set, the ground truth evidence sentences were used as positive examples. For each positive (relevant) sentence of a claim, we sampled 15 negative (irrelevant) sentences from a TF-IDF ranker: 5 from the documents that contain the positive sentences, 5 from documents that do not contain positive sentences, and another 5 such that each comes from a unique document that has not been sampled before. This sampling strategy ensures that the training set contains a diverse negative examples that better represent those encountered during inference. We find that the top negatives from TF-IDF alone often come from one or two documents and the model trained using them has lower recall than a MSMARCO fine-tuned model \citep{Dai_2019}, possibly because the negatives do not reflect those seen during inference. During training, the hardest negatives with the highest predicted relevance score are used with the positives to train the model so that the numbers of positive and negative samples match \citep{soleimani2020bert}.

\begin{table}[]
	\centering
	\caption{Distribution of numbers of entities linked to WikiData in refuted and supported claims in the original FEVER dev set.}
	\label{Entities-Statistics}
	\begin{tabular}{l l l}
		\toprule
		\# of entities & \textsc{Refuted}  & \textsc{Supported} \\
		\midrule
		$<=$ 1 entity & 4090 & 4166 \\
		$>=$ 2 entities  & 2576 & 2500 \\
		\bottomrule
	\end{tabular}
\end{table}

\begin{table}[]
	\centering
	\addtolength{\tabcolsep}{-3pt}
	\caption{Distribution of entities' relationships in refuted and supported claims in the original FEVER dev set. The entities were linked to Wikidata using the API from \citet{vanHulst:2020:REL}.}
	\label{Entities-relationship}
	\begin{tabular}{l l l}
		\toprule
		Entities' Relationship &  \textsc{Refuted} & \textsc{Supported} \\
		\midrule
		Directly Related & 571 & 998 \\
		Not Directly Related & 1928 & 1404 \\
		\bottomrule
	\end{tabular}
\end{table}

\begin{table*}[]
	\centering
	\begin{tabular}{lllllll}
		\toprule
		Model  & Baseline & REF & SUP & SR & DA  \\
		\midrule
		Recall & 0.919  & 0.864 & 0.914 & \textbf{0.936} & 0.924  \\
		\textsc{Refuted} mistakes & 330   & 271 & 542 & \textbf{241} & 308   \\
		\textsc{Supported} mistakes & 205 & 808 & \textbf{99} & {112} & 171  \\
		\bottomrule
	\end{tabular}
	\caption{Sentence selection results on the original FEVER dev set assuming perfect document retrieval system. REF model is trained only with refuted claims. SUP model is trained only with supported claims. SR model aggregates the results from SUP and REF model. DA model is trained with both original FEVER training data and synthetic false claims. See section \ref{different_models} for details.}
	\label{tab:dev-oracle-result}
\end{table*}

\begin{table}[]
	\centering
	\begin{tabular}{llll}
		\toprule
		Model  & Baseline & SR & DA\\
		\midrule
		Recall (k=5) & 0.688   & {0.767} & \textbf{0.813} \\
		\textsc{Refuted} mistakes & 282  & {187} & \textbf{141} \\
		\bottomrule
	\end{tabular}
	\caption{Sentence selection results on the 1102 artificially generated false claims from the dev set assuming perfect document retrieval system.}
	\label{tab:wikidata-aug-result}
\end{table}

\begin{table*}[]
	\centering
	\small
	\caption{An example false claim from the original FEVER dev set and corresponding top 5 evidence sentences retrieved from the baseline, DA, and SR models. The ground truth evidence sentence is in bold. Note that the title of the wikipedia article is added to the beginning of each evidence sentence for pronoun resolution.}
	\label{tab-example-pred-SR-vs.-DA-vs.-Baseline}
	\begin{tabular}{p{0.08\linewidth} | p{0.02\linewidth} | p{0.82\linewidth}}
		\toprule
		Claim &  & Johnny Galecki only acted in NBC shows.\\ 
		\midrule
		Baseline & 1 & The Tonight Show. The Tonight Show is an American late-night talk show currently broadcast from the NBC studios in Rockefeller Center in New York City (and previously from various studios in the Los Angeles region) and airing on NBC since 1954.\\
		 & 2 & The Tonight Show. It is the world 's longest-running talk show, and the longest running, regularly scheduled entertainment program in the United States.\\
		 & 3 & Blind Faith (miniseries). Blind Faith is a 1990 NBC miniseries based on the 1989 true crime book of the same name by Joe McGinniss.\\
		 & 4 & The Master Cleanse. The Master Cleanse is a 2016 American horror film written and directed by Bobby Miller.\\
		 & 5 & The Tonight Show Starring Johnny Carson. The Tonight Show Starring Johnny Carson is a talk show hosted by Johnny Carson under The Tonight Show franchise from October 1, 1962 through May 22, 1992.\\
		\midrule
		DA & 1 & The Tonight Show. The Tonight Show is an American late-night talk show currently broadcast from the NBC studios in Rockefeller Center in New York City (and previously from various studios in the Los Angeles region) and airing on NBC since 1954.\\
		 & 2 & The Tonight Show. It is the world's longest-running talk show, and the longest running, regularly scheduled entertainment program in the United States.\\
		 & 3 & AFL on NBC. The AFL on NBC is the branding used for broadcasts of Arena Football League (AFL) games produced by NBC Sports, the sports division of the NBC television network in the United States, that aired from the 2003 to 2006 seasons.\\
		 & 4 & \textbf{Johnny Galecki. He is known for playing David Healy in the ABC sitcom Roseanne from 1992 -- 1997 and Dr. Leonard Hofstadter in the CBS sitcom The Big Bang Theory since 2007.}\\
		 & 5 & Blind Faith (miniseries). Blind Faith is a 1990 NBC miniseries based on the 1989 true crime book of the same name by Joe McGinniss.\\
		\midrule
		SR & 1 & Johnny Galecki. Galecki also appeared in the films National Lampoon's Christmas Vacation (1989), Prancer (1989), Suicide Kings (1997), I Know What You Did Last Summer (1997), Bookies (2003), and In Time (2011).\\
		 & 2 & The Tonight Show. The Tonight Show is an American late-night talk show currently broadcast from the NBC studios in Rockefeller Center in New York City (and previously from various studios in the Los Angeles region) and airing on NBC since 1954.\\
		 & 3 & The Master Cleanse. The Master Cleanse is a 2016 American horror film written and directed by Bobby Miller.\\
		 & 4 & \textbf{Johnny Galecki. He is known for playing David Healy in the ABC sitcom Roseanne from 1992 -- 1997 and Dr. Leonard Hofstadter in the CBS sitcom The Big Bang Theory since 2007.}\\
		 & 5 & NBC. The National Broadcasting Company (NBC) is an American English language commercial broadcast television network that is a flagship property of NBCUniversal, a subsidiary of Comcast.\\
		\bottomrule
	\end{tabular}
\end{table*}

\begin{table*}[]
	\centering
	\begin{tabular}{lllll}
		\toprule
		Model  & Baseline & SR & DA  & \citet{soleimani2020bert}*\\
		\midrule
		Recall (k=5) & 0.868  & \textbf{0.882} &  0.8721  & 0.752 \\
		FEVER Score & 0.656   & \textbf{0.660} & 0.658  & 0.611 \\
		Label accuracy & 0.705 & \textbf{0.707} & 0.707 & 0.678 \\
		\bottomrule
	\end{tabular}
	\caption{FEVER results on the blind test set. * indicates the results are from our implementation. }
	\label{tab:blind-test-result}
\end{table*}

\subsection{Natural Language Inference}
We adopted a similar method from \citet{soleimani2020bert} for natural language inference to have a complete FEVER pipeline to evaluate our sentence selection methods on the blind test set. A pre-trained 3-class classification BERT model is fine-tuned to classify the relationship between an evidence sentence and a claim as \textit{\textsc{Refuted}}, \textit{\textsc{Supported}}, or \textit{\textsc{NotEnoughInfo}} using cross-entropy loss. During inference, we classify the relationship between the top 5 evidence sentences with the claim separately and aggregate the results using majority vote. If there is a tie, we break it using the following sequence: \textit{\textsc{NotEnoughInfo}} to \textit{\textsc{Supported}} to \textit{\textsc{Refuted}}.

\section{Results}
\label{results}

\textbf{False Claims are More Likely to Have Unrelated Entities than True Claims.} 
For each refuted or supported claim in the original FEVER dev set, we programmatically linked its entities to WikiData using the API from \citet{vanHulst:2020:REL}. If two or more entities were identified, we check whether there is any relationship between any 2 entities using WikiData SPARQL. If any 2 entities are related in any relationship, we say they are directly related. Otherwise, they are not directly related. Table \ref{Entities-relationship} shows that the refuted claims are significantly more likely to have irrelevant entities than supported claims (${\chi}^2$ test, $p < 0.01$, ${\chi}^2(1) = 195.91$), even though refuted and supported claims have similar distributions in terms of number of entities (${\chi}^2$ test, $p > 0.1$, ${\chi}^2(1) = 1.79$, Table \ref{Entities-Statistics}). This result shows that refuted claims are more likely to contain unrelated or irrelevant entities that may be distracting to a retrieval model than supported claims. \cmmnt{We used the dev set because we do not know whether a claim is supported or refuted in the test set whereas the training set has unequal numbers of refuted and claimed claims.}

\textbf{Baseline BERT-based Retrieval Model is Not Robust to Irrelevant Entities in Claims.}
To compare how various models differ in selecting refuting and supporting sentences without being biased by the document retrieval results (Table \ref{table-document-retrieval}), the ground truth documents are appended to those from the document retrieval stage. This method simulates a realistic scenario where the sentence selection model has to retrieve the relevant sentences among a large pool of irrelevant sentences. Table \ref{tab:dev-oracle-result} shows that the baseline model made more refuted mistakes than supported mistakes. This result, together with the fact that false claims are more likely to have irrelevant entities than true claims, showing that the baseline model is not robust to irrelevant entities in the false claims. To further test this hypothesis, we evaluated the recall of the baseline model in retrieving evidence for the artificially generated false claims with irrelevant entities from the dev set. Table \ref{tab:wikidata-aug-result} and \ref{tab:dev-oracle-result} shows that the baseline model had lower recall on the adversarial false claims with irrelevant entities than on the original claims, which supports our hypothesis that the baseline model is not robust to irrelevant entities in false claims.

\textbf{Improve the Model Robustness to Irrelevant Entities in False Claims.}
 Table \ref{tab:dev-oracle-result} shows that SUP and REF model made fewer mistakes than the baseline model in retrieving supporting and refuting evidence, respectively, suggesting that recall may be improved by retrieving refuting and supporting evidence separately and then aggregating. Indeed, on the original dev set, the highest recall is achieved by the SR model, which combines the SUP and REF results. The SR model also made fewer mistakes in retrieving refuting evidence than the baseline model (Table \ref{tab:dev-oracle-result}). When tested on the artificially generated false claims with irrelevant entities, the SR model has higher recall than baseline model (Table \ref{tab:wikidata-aug-result}). These results together show that SR model is more robust to irrelevant entities in the false claims than the baseline model (see Table \ref{tab-example-pred-SR-vs.-DA-vs.-Baseline} for an example). Trained with additional synthetic refuted claims, the DA model achieved higher recall than baseline model and made fewer refuted mistakes on the original dev set (Table \ref{tab:dev-oracle-result}). When tested on the artificially generated false claims with irrelevant entities, the DA model achieves the highest recall (Table \ref{tab:wikidata-aug-result}). These results show that data augmentation is another effective method to improve the retrieval model's robustness to irrelevant entities in false claims (Table \ref{tab-example-pred-SR-vs.-DA-vs.-Baseline}). Note that our current data augmentation method is far from ideal because the irrelevant entity is always the 2nd entity in the artificially generated false claims. Consequently, the DA model has lower recall than the SR model on the original dev and blind test set (Table \ref{tab:dev-oracle-result} \& \ref{tab:blind-test-result}), even though the DA model has higher recall than the SR/baseline model for the artificially generated adversarial false claims with irrelevant entities (Table \ref{tab:wikidata-aug-result}).
 

\textbf{Full FEVER Pipeline.}
Table \ref{tab:blind-test-result} shows that our robust models (SR \& DA) achieves similar or higher recall and FEVER score on the blind test set compared with the baseline, showing that our methods generalize to the blind test set. Note that the gain in recall and FEVER score is smaller in the blind test set than those seen in the validation set. It may be because the document retrieval model is not robust to false claims with irrelevant entities and sentence selection model cannot recover from these errors when evaluated on the blind test set whereas we assumed a perfect document retrieval model for the validation set. In addition, DA \& SR models perform significantly better on the generated false claims with irrelevant entities than the baseline model, suggesting DA \& SR model will be at least more robust to these examples than the baseline model. Note that the result from \citet{soleimani2020bert} in our implementation is lower than that reported in their paper. This difference is probably due to different pre-processing steps because \citet{soleimani2020bert} only open-sourced the model training scripts but not the preprocessing scripts that generate the training data.

\section{Conclusion \& Future Directions}
Our results, together with those from \citet{Yoneda2018UCLMR} and \citet{Chakrabarty2018RobustDR}, show that evidence retrieval is more challenging for false claims than for true claims. One possible reason is that false claims more often contain irrelevant entities than true claims and the baseline model is not robust to irrelevant entities in a claim. When the training data is augmented with additional synthetic false claims with irrelevant entities, the same model achieves higher evidence recall than when there are no augmented data. When separate models are trained to retrieve refuting and supporting evidence, the model for refuting evidence can learn to ignore irrelevant entities whereas the model for supporting evidence can learn to retrieve evidence that has the largest lexical and semantic overlap with the claim. After aggregating the results from the two models, we achieved the highest sentence selection recall. In the current paper, we only investigated how to improve the robustness of sentence selection model. In the future, we will look into improving the robustness of the document retrieval and natural language inference model. For example, instead of always replacing the 2nd entity in a claim, we can dynamically replace the entity while considering the relationship between entities. If synthetic claims of higher quality can be obtained, the synthetic data can be used to train the document retrieval and NLI model to further increase the model robustness. Another future work is to investigate whether the state-of-the-art models (e.g., \citet{hidey2020deseption}) are robust to irrelevant entities in the false claims, which can potentially reveal additional ways to increase the model robustness. In the end, we suggest that future research to report separate metrics for false and true claims. In the real world, false claims but not true claims are harmful to the society. Therefore, it is important to know our models retrieve refuting evidence as well as (or better than) supporting evidence.



\bibliography{acl2019}
\bibliographystyle{acl_natbib}

\end{document}